\documentclass[aps,showpacs,showkeys,superscriptaddress]{revtex4-1}

\usepackage{graphicx}

\begin{document}

\title{The string prediction models as an invariants of time series in forex market}

\author{R. Pincak}\email{pincak@sors.com}
\author{M. Repasan}\email{repasan@sors.com}
\affiliation{SORS Research a.s, 040 01 Kosice, Slovak Republic}


\date{\today}

\begin{abstract}

In this paper we apply a new approach of the string theory to the
real financial market. It is direct extension and application of the
work~\cite{Pincak} into prediction of prices. The models are
constructed with an idea of prediction models based on the string
invariants (PMBSI). The performance of PMBSI is compared to support
vector machines (SVM) and artificial neural networks (ANN) on an
artificial and a financial time series. Brief overview of the
results and analysis is given. The first model is based on the
correlation function as invariant and the second one is an
application based on the deviations from the closed string/pattern
form (PMBCS). We found the difference between these two approaches.
The first model cannot predict the behavior of the forex market with
good efficiency in comparison with the second one which is, in
addition, able to make relevant profit per year.

\end{abstract}

\maketitle

\section{Introduction}

The time-series forecasting is a scientific field under continuous
active development covering an extensive range of methods.
Traditionally, linear methods and models are used. Despite their
simplicity, linear methods often work well and may well provide an
adequate approximation for the task at hand and are mathematically
and practically convenient. However, the real life generating
processes are often non-linear. This is particularly true for
financial time series forecasting. Therefore the use of non-linear
models is promising. Many observed financial time series exhibit
features which cannot be explained by a linear model.

There are plenty of non-linear forecast models based on different
approaches (e.g. GARCH \cite{GARCH}, ARCH \cite{ARCH}, ARMA
\cite{ARMA}, ARIMA \cite{ARIMA} etc) used in financial time series
forecasting. Currently, perhaps the most frequently used methods are based on
Artificial Neural Networks (ANN, covers a wide range of methods) and
Support Vector Machines (SVM). A number of research articles
compares ANN and SVM to each other and to other more traditional
non-linear statistical methods. Tay and Cao (\cite{Tay2001})
examined the feasibility of SVM in financial time series forecasting
and compared it to a multilayer Back Propagation Neural Network
(BPNN). They showed that SVM outperforms the BP neural network.
Kamruzzaman and Sarker \cite{Kamruzzaman2003} modeled and predicted
currency exchange rates using three ANN based models and a
comparison was made with ARIMA model. The results showed that all
the ANN based models outperform ARIMA model. Chen et al.
\cite{Chen2006} compared SVM and BPNN taking auto-regressive model
as a benchmark in forecasting the six major Asian stock markets.
Again, both the SVM and BPNN outperformed the traditional models.

While the traditional ANN implements the empirical risk minimization
principle, SVM implements the structural risk minimization
(\cite{Vapnik1995}). Structural risk minimization is an inductive
principle for model selection used for learning from finite training
data sets. It describes a general model of capacity control and
provides a trade-off between hypothesis space complexity and the
quality of fitting the training data (empirical error). For this
reason SVM is often chosen as a benchmark to compare other
non-linear models to. Also, there is a growing number of novel and
hybrid approaches, combining the advantages of various methods using
for example evolutionary optimization, methods of computational
geometry and other techniques (e.g. \cite{Bundzel2006},
\cite{BKF2006}).

In this paper we apply the string model and approaches described
in~\cite{Pincak} to real finance forex market. This is an extension
of the previous work~\cite{Pincak} into the real finance market. We
derive two models for predictions of EUR/USD prices on the forex
market. This is the first attempt for real application of the string
theory in the field of finance, and not only in high energy physics,
where it is established very well. Firstly we described briefly some
connections between these different fields of research.

We would like to transfer modern physics ideas into neighboring
field called econophysics.  The physical statistical viewpoint has
proved to be fruitful, namely in the description of systems where
many-body effects dominate. However, standard, accepted by
physicists, bottom-up approaches are cumbersome or outright
impossible to follow the behavior of the complex economic systems,
where autonomous models encounter intrinsic variability.

Modern digital economy is founded on data. Our primary motivation
comes from the actual physical concepts~\cite{Mahon2009,Zwie2009};
however, our realization differs from the original attempts in
various significant details. Similarly as with most scientific
problems, the representation of data is the key to efficient and
effective solutions. The string theory development over the past 25
years has achieved a high degree of popularity among
physicists~\cite{Polchinski}.

The underlying link between our approach and the string theory may
be seen in switching from a local to a non-local form of
data description. This line passes from the single price to the
multivalued collection, especially the string of prices from the
temporal neighborhood, which we term here as string map. It is the
relationship between more intuitive geometric methods and financial
data. Here we work on the concept that is based on projection data
into higher dimensional vectors in the sense of the
works~\cite{Grass1983,Ding2010}.

The present work exploits time-series which can build the family of
the string-motivated models of boundary-respecting maps. The purpose
of the present data-driven study is to develop statistical
techniques for the analysis of these objects and moreover for the
utilization of such string models onto the forex market. Both of the
string prediction models in this paper are built on the physical
principle of the invariant in time series of forex market. Founding
of a stationary states in time series of market was studied
in~\cite{Stanley}

\section{Definition of the strings}

By applying standard methodologies of detrending we suggest to
convert original series of the quotations of the  mean currency
exchange rate $p(\tau)$ onto a series of returns defined by
\begin{equation}
\frac{p(\tau+h)  -  p(\tau)}{p(\tau+h)}\,,
\end{equation}
where $h$ denotes a tick lag between currency quotes $p(\tau)$ and
$p(\tau+h)$, $\tau$ is the index of the quote. The mean
$p(\tau)=(p_{\rm ask}(\tau)+ p_{\rm bid}(\tau))/2$ is calculated
from $p_{\rm ask}(\tau)$ and $p_{\rm bid}(\tau)$.

In the spirit of the string theory it would be better to start with
the 1-end-point open string map
\begin{equation}
P^{(1)}(\tau,h) = \frac{p(\tau+h)  - p(\tau)}{p(\tau+ h)}\,,
\qquad h \in <0,l_{\rm s}>
\label{eq:string1}
\end{equation}
where the superscript $(1)$ refers to the number of endpoints.

The variable $h$ may be interpreted as a variable which extends
along the extra dimension limited by the string size $l_{\rm s}$.
For the natural definitions of the string must be fulfilled the
boundary condition
\begin{equation}
P^{(1)}(\tau,0)= 0\,,
\end{equation}
which holds for any tick coordinate $\tau$. We want to highlight the
effects of rare events. For this purpose, we introduce a
power-law Q-deformed model
\begin{equation}
P^{(1)}_q(\tau,h) =
\left(1-\left[\frac{p(\tau)}{p(\tau+h)}\right]^{Q}\right),\qquad
Q>0\,.\label{eq:stringq}
\end{equation}
The 1-end-point string has defined the origin, but it reflects the
linear trend in $p(.)$ at the scale $l_{\rm s}$. Therefore, the
1-end-point string map $P^{(1)}_q(.)$ may be understood as a
Q-deformed generalization of the {\em currency returns}.

The situation with a long-term trend is partially corrected by
fixing $P^{(2)}_q(\tau,h)$ at $h=l_{\rm s}$. The open string with
two end points is introduced via the nonlinear map which combines
information about trends of $p$ at two sequential segments
\begin{equation}
P^{(2)}_q(\tau,h)
=\left(1-\left[\frac{p(\tau)}{p(\tau+h)}\right]^{Q}\right)\left(1-\left[\frac{p(\tau+h)}{p(\tau+l_{\rm
s})}\right]^{Q}\right)
 \,,\qquad  h \in  <0,l_{\rm s}>\,.
\label{eq:Pq2p}
\end{equation}
The map is suggested to include boundary conditions of {\em Dirichlet type}
\begin{equation}
P^{(2)}_q(\tau,0)= P_q(\tau,l_{\rm s})=0\,,\qquad
\mbox{at all ticks}\,\,\, \tau\,.
\label{eq:Dir}
\end{equation}
In particular, the sign of $P^{(2)}_q(\tau,h)$ comprises information
about the behavior differences of $p(.)$ at three quotes $(\tau,\,
\tau+h,\, \tau+l_{\rm s})$.

Now we define partially compactified strings. In the frame of the
string theory, compactification attempts to ensure compatibility
of the universe based on the four observable dimensions with
twenty-six dimensions found in the theoretical model systems.  From
the standpoint of the problems considered here,compactification
may be viewed as an act of information reduction of the original
signal data, which makes the transformed signal periodic. Of course,
it is not very favorable to close strings by the complete
periodization of real input signals. Partial closure would be more
interesting. This uses pre-mapping
\begin{equation}
\tilde{p}(\tau) = \frac{1}{N_{\rm m}} \sum_{m=0}^{N_{\rm m}-1}
p(\tau +l_{\rm s} m)\,, \label{eq:compact}
\end{equation}
where the input of any open string (see e.g. Eq.(\ref{eq:string1}),
Eq.(\ref{eq:Pq2p})) is made up partially compact.

Thus, data from the interval $<\tau, \tau+l_{\rm s} ( N_{\rm m}-1 )
>$ are being pressed to occupy "little space" $h\in <0, l_{\rm s}>$.
We see that as $N_{\rm m}$ increases, deviations of $\tilde{p}$ from
the periodic signal become less pronounced. The corresponding
statistical characteristics of all the strings and brane described
above were displayed in detail in~\cite{Pincak}. The prediction
models presented in the paper were tested on the tick by tick one
year data of EUR/USD major currency pair from ICAP market maker.
More precisely, we selected period from October 2009 to September
2010.

\section{Correlation function as invariant}\label{sec:polar}
The meaning of invariant is that something does not change under
transformation, such as some equations from one reference frame
to another. We want to extend this idea also on the finance market, find some invariants in the finance data and utilize this
as prediction for the following prices. Unfortunately this model is
able to define only one step prediction, see definition below.

We suppose the invariant in a form of correlation function
\begin{eqnarray}
C_{(t,l_0)} &=& \sum_{h=l_0}^{h=l} w_h \left(1 -
\frac{p_{t-h}}{p_{t-1-h}}\right) \left(1 -
\frac{p_{t-1-h}}{p_{t-2-h}}\right),
\end{eqnarray}
with
\begin{eqnarray}
w_h &=& \frac{e^{-h/\lambda}}{\sum_{h'=0}^l e^{-h'/\lambda}},
\end{eqnarray}
including dependence on the time scale parameters $l, l_0$ and
$\lambda$. The relative weights satisfy automatically $\sum_{h=0}^l
w_h = 1$. From the condition of the invariance at the step $(t+1)$
\begin{equation}
C_{(t,0)} = C_{(t+1,0)} \simeq \hat{C}_{(t+1,0)}
\end{equation}
where $\hat{C}$ only symbolically meaning some stationarity in the
next step of correlation function and we finally obtain the
prediction
\begin{equation}
\hat{p}_{t+1} =  p_t \, \left( 1 + \frac{C_{(t+1,1)}  -  C_{(t,0)}
}{w_0 \left( 1 -  \frac{p_t}{p_{t-1}}  \right) } \, \right),
\end{equation}
valid for $p_t\neq p_{t-1}$. These are general definitions for the
correlation invariants.

\subsection{Prediction model based on the string invariants (PMBSI)}\label{sec:polar}

Now we want to take the above-mentioned ideas onto the string maps
of finance data. We would like to utilize the power of the nonlinear
string maps of finance data and establish some prediction models to
predict the behavior of market similarly as in the
works~\cite{Christiansen,Chang,Wolfers}. We suggest the method where
one string is continuously deformed into the other. We analyze
1-end-point and 2-end-point mixed string models. The family of
invariants is written using parametrization
\begin{eqnarray}
C(\tau,\Lambda)  &=&  (1-\eta_1) (1-\eta_2) \sum_{h=0}^{\Lambda}
W(h)
\\
&\times & \left( 1- \left[\frac{p(\tau)}{p(\tau+h)}\right]^Q
\right) \, \left( 1-\left[\frac{p(\tau+h)}{p(\tau+l_s)}\right]^Q
\right) \nonumber
\\
&+& \eta_1 (1-\eta_2) \sum_{h=0}^{\Lambda} W(h)\, \left(   1-
\left[\frac{p(\tau)}{p(\tau+h)}\right]^Q   \right)\, \nonumber
\\
&+& \eta_2 \sum_{h=0}^{\Lambda} W(h)\, \left(   1-
\left[\frac{p(\tau+h)}{p(\tau+l_s)}\right]^Q \right)\,,
\end{eqnarray}
where $\eta_1\in (-1,1)$, $\eta_2\in (-1,1)$ are variables
(variables which we may be called homotopy parameters),  $Q$ is a
real valued parameter, and the weight $W(h)$ is chosen in the
bimodal single parameter form
\begin{eqnarray}
W(h)= \left\{
\begin{array}{ll}
1 - W_0 \,,  &   h   \leq  l_s/2\,,
\\
W_0     \,,  &   h    >    l_s/2\,.
\end{array}
\right.
\end{eqnarray}
We plan to express $p(\tau+l_s)$ in terms of the auxilliary
variables
\begin{eqnarray}
A_1(\Lambda) &=&  ( 1-\eta_1) ( 1-\eta_2)  \sum_{h=0}^{\Lambda} W(h)
\left(1-\left[\frac{p(\tau)}{p(\tau+h)} \right]^Q\right)\,,
\\
A_2(\Lambda) &=& -( 1-\eta_1) ( 1-\eta_2)  \sum_{h=0}^{\Lambda} W(h)
\left(1-\left[\frac{p(\tau)}{p(\tau+h)} \right]^Q\right)
p^Q(\tau+h)\,,
\\
A_3(\Lambda) &=& \eta_1 (1-\eta_2) \sum_{h=0}^{\Lambda} W(h)
\left(1-\left[\frac{p(\tau)}{p(\tau+h)} \right]^Q\right)\,,
\\
A_4(\Lambda) &=& \eta_2 \sum_{h=0}^{\Lambda} W(h) \,,
\\
A_5(\Lambda) &=& - \eta_2  \sum_{h=0}^{\Lambda} W(h)\,
p^Q(\tau+h)\,.
\end{eqnarray}
Thus the expected prediction form reads
\begin{equation}
\hat{p}(\tau_0+l_{\rm pr})= \left[\frac{ A_2(\Lambda) + A_5(\Lambda)
}{ C(\tau_0-l_{\rm s},\Lambda) - A_1(\Lambda) -A_3(\Lambda) -
A_4(\Lambda) } \right]^{1/Q}\,,
\end{equation}
where we use the notation $\tau = \tau_0  +  l_{\rm pr} - l_s$. The
derivation is based on the invariance
\begin{equation}
C(\tau,l_s - l_{\rm pr}) =  C(\tau-l_{\rm pr}, l_s-l_{\rm pr})\,,
\quad \Lambda = l_s-l_{\rm pr}\,,
\end{equation}
where $l_{\rm pr}$ denotes the prediction scale.

The model was tested for various sets of parameters $l_s$, $l_{\rm
pr}$, $\eta_1$, $\eta_2$, $Q$ and the new parameter $\epsilon$ which
is defined as
\begin{equation}
\epsilon = \left|C(\tau, l_s - l_{\rm pr}) -
C(\tau-l_{\rm pr}, l_s-l_{\rm pr})\right|
\end{equation}
and describes the level of invariance in real data. The best
prediction (the best means that the model has the best ability to
estimate the right price) is obtained by using the following values of
parameters
\begin{eqnarray}
l_s &=& 900, \nonumber \\
l_{\rm pr} &=& 1, \nonumber \\
\eta_1 &=& 0, \nonumber \\
\eta_2 &=& 0, \nonumber \\
Q &=& 6, \nonumber \\
\epsilon &=& 10^{-10}.
\end{eqnarray}
The graphical descriptions of prediction behavior of the model
with and without transaction costs on the EUR/USD currency rate of
forex market are described in Figs 1-4. During one year period the
model lost around $20\%$ of initial money. It executed $1983$ trades
(Fig 1) where only $10$ were suggested by the model (and earned
money) and the rest of them were random (which can be clearly seen
in Figs 3,4). The problem of this model is its prediction length
(the parameter $l_{\rm pr}$), in this case it is one tick ahead. The
price was predicted correctly in $48.57\%$ of all cases ($16201$ in
one year) and from these $48.57\%$ or numerally $7869$ cases only
$0.13\%$ or numerally $10$ were suitable for trading. This small
percentage is caused by the fact that the price does not change too
often one tick ahead. One could try to raise the prediction length
to find more suitable cases for trading. This is only partly
successful because the rising parameter $l_{\rm pr}$ induces a loss
of the prediction strength of the model. For example when $l_{\rm
pr} = 2$ (two ticks ahead) prediction strength decreases from around
$50\%$ to $15\%$.

\begin{figure}
 \begin{center}
 \includegraphics[height=7cm]{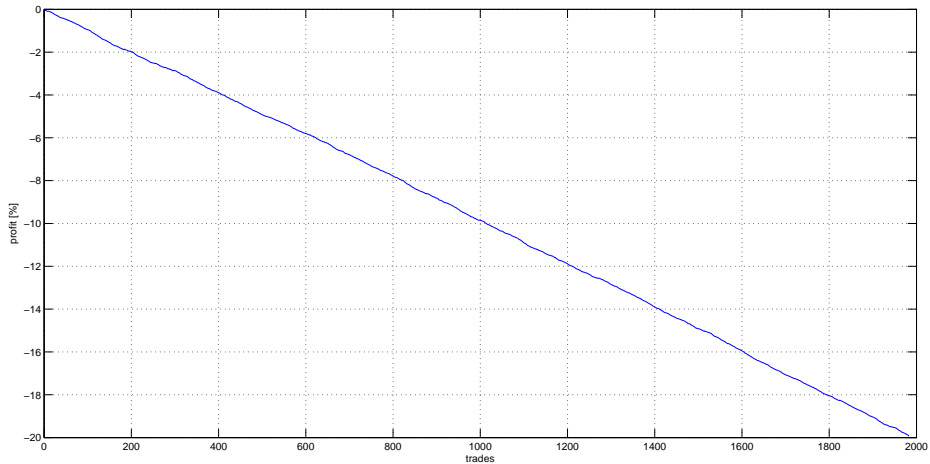}
 \caption{The profit of the model on the EUR/USD currency rate with transaction costs included dependence on trades for one year period.}
 \end{center}
 \end{figure}

\begin{figure}
 \begin{center}
 \includegraphics[height=7cm]{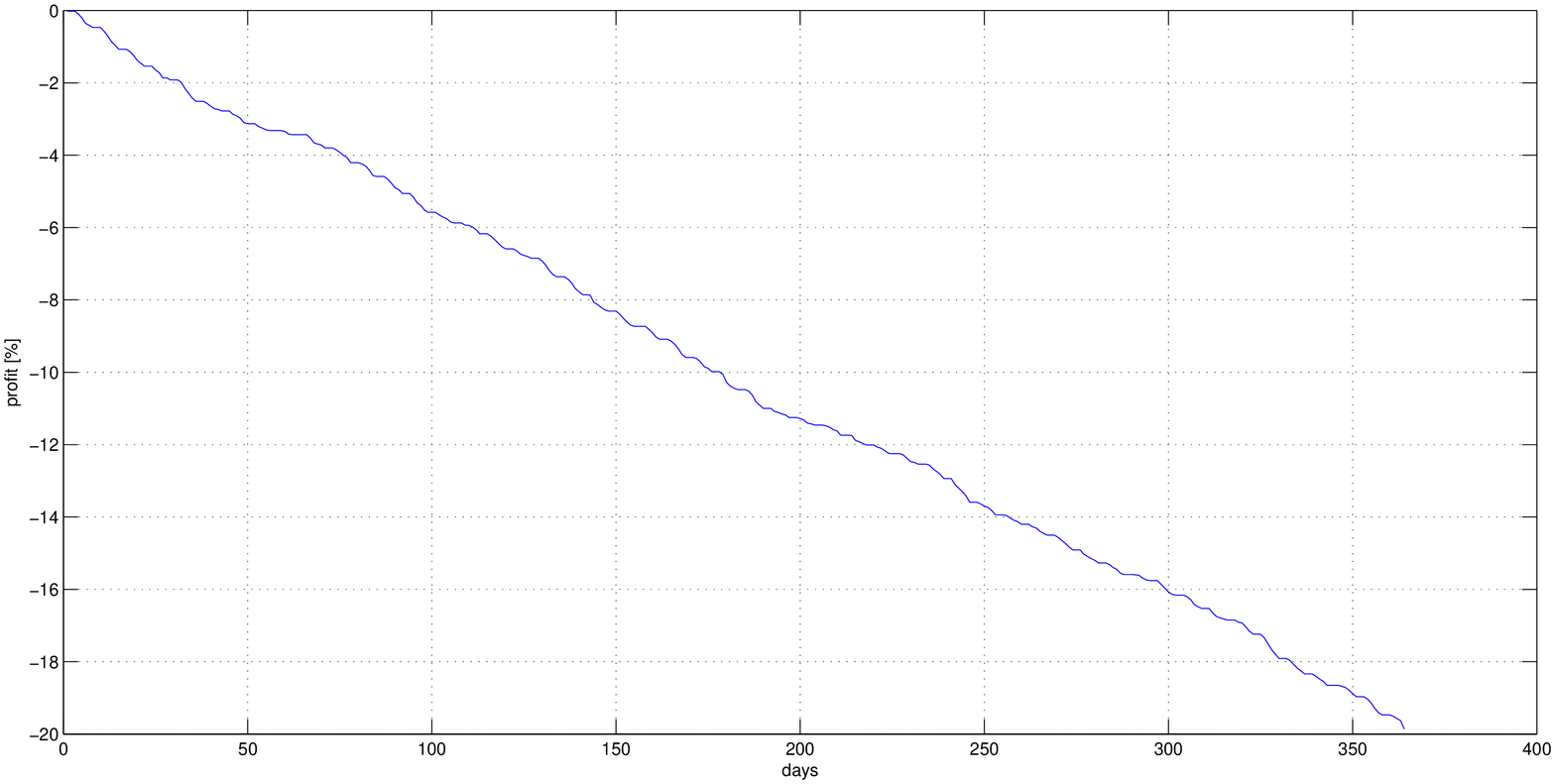}
 \caption{The profit of the model on the EUR/USD currency rate with transaction costs included dependence on days for one year period.}
 \end{center}
 \end{figure}

\begin{figure}
 \begin{center}
 \includegraphics[height=7cm]{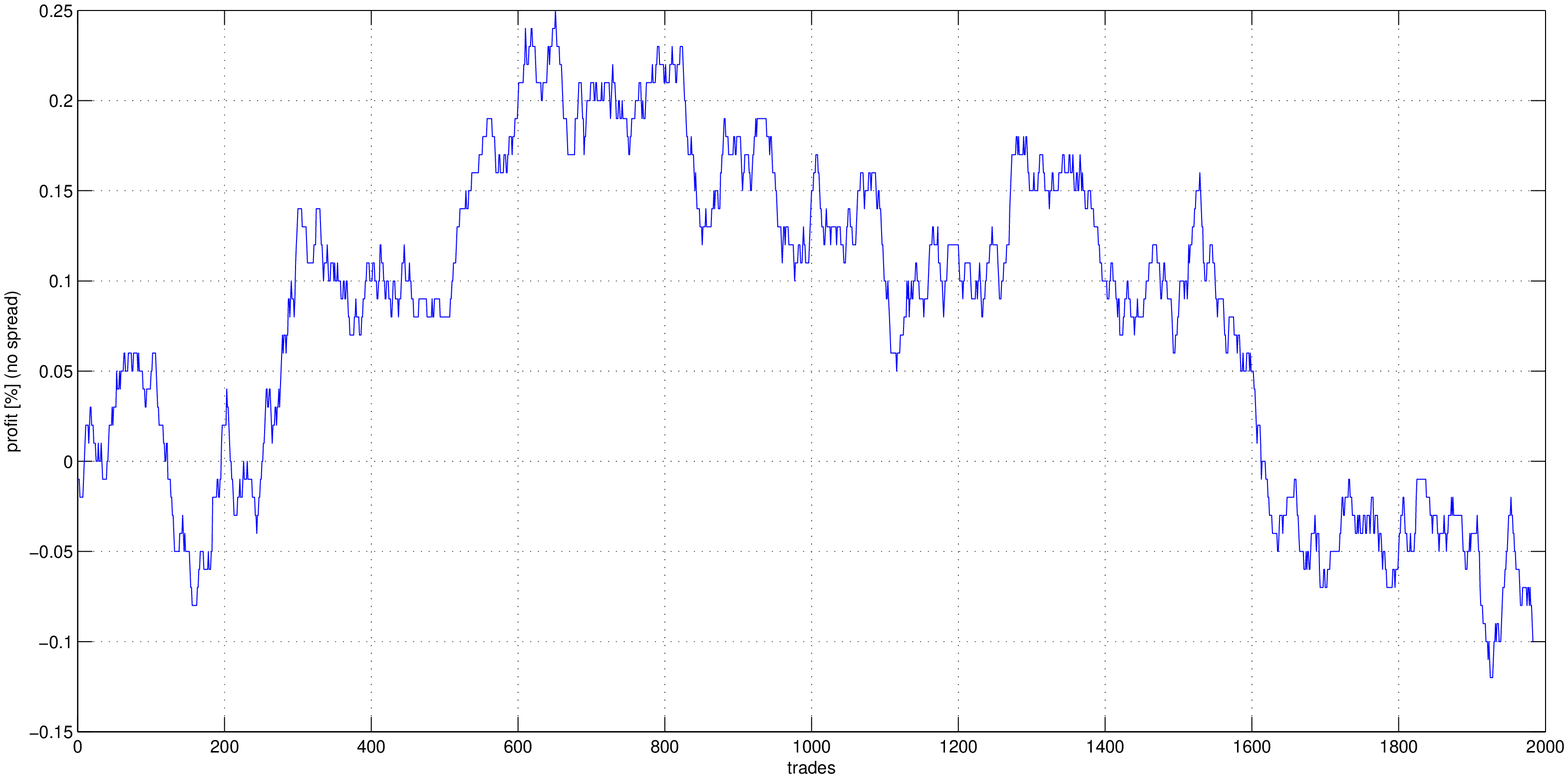}
 \caption{The profit of the model on the EUR/USD currency rate without transaction costs included dependence on trades for one year period.}
 \end{center}
 \end{figure}

\begin{figure}
 \begin{center}
 \includegraphics[height=7cm]{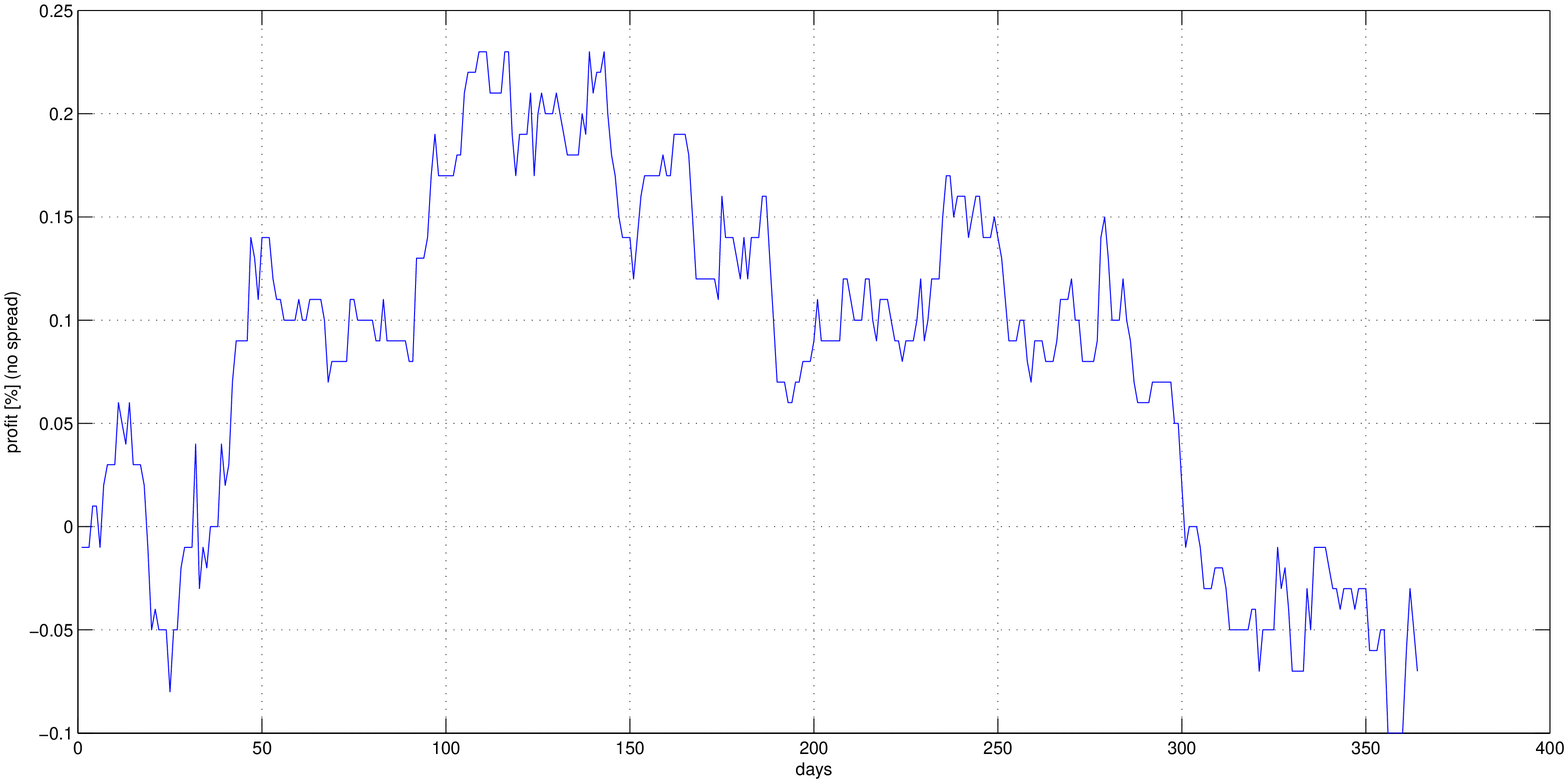}
 \caption{The profit of the model on the EUR/USD currency rate without transaction costs included dependence on days for one year period.}
 \end{center}
 \end{figure}

The problem is that the invariant equation 10 is fulfilled only on
the very short period of the time series due to the very chaotic nature
of financial data behaviour. Therefore the PMBSI is effective
only on the one step prediction where there is very low probability
that time series change significantly. The situation, however, is different
for more steps prediction where there is, on the contrary, very high
probability of big changes in time series to occur, and the following predictions
have rather small efficiency in such cases. The only way how
to establish better prediction also for more steps prediction is to choose the right weights Eq.9. The right and optimized weights should
considerably extend the interval where equation 10 is fulfilled.
Therefore it is also our task in the future work.

\subsection{Experimental Setup}
The experiments were performed on two time-series. The first series
represented artificial data namely a single period of a sinusoid
sampled by 51 regularly spaced samples. The second time series
represented proprietary financial data sampled daily over the period
of 1295 days. The performance of PMBSI was compared to SVM and to
naive forecast. There were two error measures used, mean absolute
error (MAE) and symmetric mean absolute percentage error (SMAPE)
defined as follows:
\begin{eqnarray}
MAE &=& \frac{1}{n}\sum_{t=1}^{n}|A_t - F_t|\,,
\\
SMAPE &=& \frac{100}{n}\sum_{t=1}^{n}\frac{|A_t - F_t|}{0.5(|A_t| +
|F_t|)}\,,
\end{eqnarray}
where $n$ is the number of samples, $A_t$ is the actual value and
$F_t$ is the forecast value. Each time-series was divided into three
subsets: training, evaluation and validation data. The time ordering
of the data was maintained; the least recent data were used for
training, the more recent data were used to evaluate the performance
of the particular model with the given parameters' setting. The best
performing model on the evaluation set (in terms of MAE) was chosen
and made forecast for the validation data (the most recent) that
were never used in the model optimization process. Experimental
results on the evaluation and validation data are presented below.
The parameters of the models were optimized by trying all
combinations of parameters sampled from given ranges with a
sufficient sampling rate. Naturally, this process is slow but it
enabled to get an image of the shape of the error surface
corresponding to the given settings of parameters and ensured that
local minima are explored. The above approach was used for both,
PMBSI and SVM. The SVM models were constructed so that the present
value and a certain number of the consecutive past values comprised
the input to the model. The input vector corresponds to what will be
referred to here as the {\it time window} with the length $l_{tw}$
(representing the equivalent of the length of the string map $l_s$
by PMBSI).

\section{Comparison}
There was a preliminary experimental analysis of the PMBSI method
performed. The goal was to evaluate the prediction accuracy,
generalization performance, convenience of the method in terms
of the operators effort needed to prepare a working model,
computational time and other aspects of the PMBSI method that may
have become obvious during the practical deployment. SVM was chosen
as a benchmark. The experimental data comprised two sets: artificial
data (a single period of a sinusoid) and real world data (financial,
price development). We will provide a brief conclusion of the
analysis here. Each time series was divided into three subsets for
training, testing and validation. The results were calculated on the
validation sets that have been entirely absent in the process of optimization of parameters.

\begin{figure}[htbp]
\begin{center}
 \includegraphics[height=10cm]{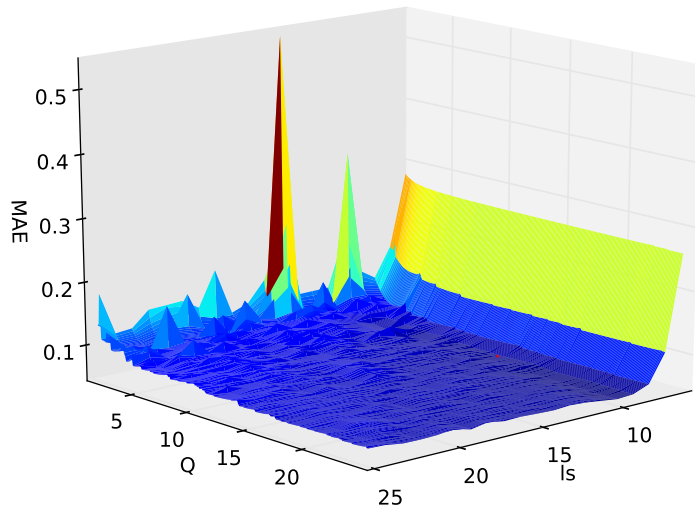}\caption{MAE
corresponding to various settings of $ls$ and $Q$ on the financial
data. The red dot is the global minimum of MAE.}
\label{ErrorSurface}
\end{center}
\end{figure}

PMBSI predictor does not undergo a training process that is typical
for ANN and SVM where a number of free parameters must be set
(synaptic weights by ANN, $\alpha$ coefficients by SVM). PMBSI
features a similar set of weights ($W$) but often very small and
calculated analytically. The parameters to be optimized are only
four: $ls$, $Q$, $\eta_1$, $\eta_2$. This, clearly, is an advantage.
On the other hand the optimal setting of the parameters is not easy
to be found as there are many local minima on the error surface. In
this analysis the optimal setting was found by testing of all
combinations of parameters from given ranges. Fig.
\ref{ErrorSurface} shows the Mean Absolute Error (MAE) of the
5-steps ahead forecast of the financial time series corresponding to
various settings of $ls$ and $Q$ ($\eta_1,\eta_2 = 0$). But the
figure makes also obvious that PMBSI's performance  is approximately
the same for a wide range of settings on this data.

\begin{figure}[htbp]
\begin{center}
 \includegraphics[height=10cm]{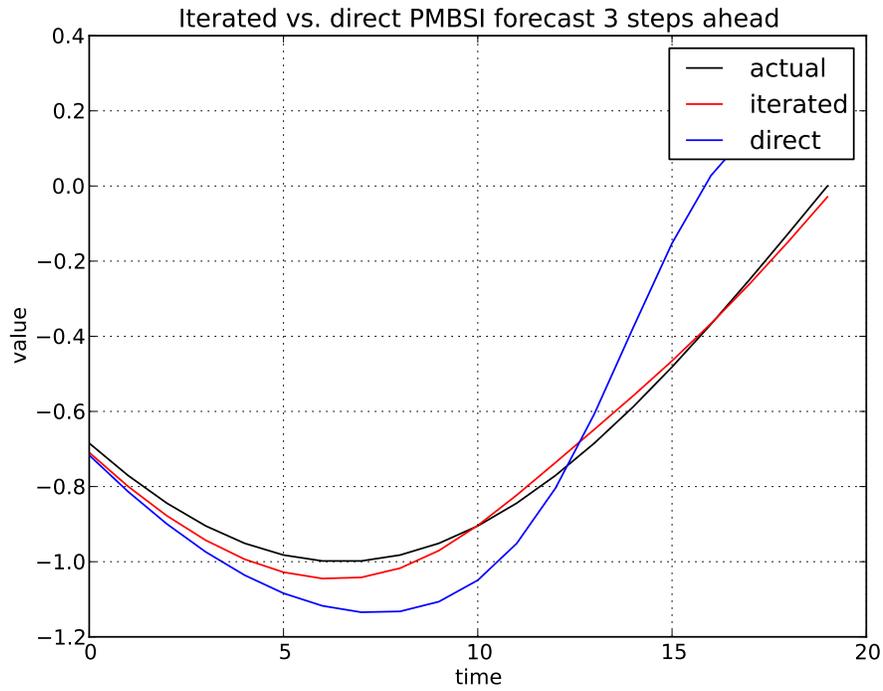}
\caption{Iterated and direct prediction using PMBSI on artificial
data.} \label{IvsD}
\end{center}
\end{figure}

For PMBSI to work the elements of time series must be non zero
otherwise the method will return {\it not a number} forecasts only.
The input time series must be then modified by adding a constant and
the forecast by subtracting the same constant. Even so the algorithm
returned a {\it not a number} forecast in app. 20\% of the cases on
the financial data. In such cases the last valid forecast was used.
Due to reasons that are presently being evaluated the accuracy of
PMBSI is matching and even outperforming SVM for a single step
predictions but rapidly deteriorates for predictions of more steps
ahead. Iterated prediction of several steps ahead using the single
step PMBSI predictor improves the accuracy significantly. The
sinusoid used for experiments was sampled by 51 points, the positive
part of the wave was used for optimization of the parameters and the
rest for validation (app. 50-50 division). Fig. \ref{IvsD} shows the
comparison of iterated versus the direct prediction using PMBSI.
Table \ref{res_artificial} shows the experimental results. The
results of the best performing models are highlighted.

\begin{table}
\begin{center}
\begin{tabular}[c]{|c|c|c|c|c|}
\hline
Method & $l_{pr}$ & MAE & MAE & SMAPE \\
 & & eval & valid & valid \\ \hline \hline
{PMBSI} & 1 & 0.000973 &    {\bf 0.002968} & {\bf 8.838798}  \\
\cline{2-5} & 2 &  0.006947 &   0.034032 & 14.745538
\\ \cline{2-5} & 3 &  0.015995 & 0.161837  & 54.303315
\\ \hline \hline {Iterated PMBSI} & 1 & - & - & -
\\ \cline{2-5} & 2 & 0.003436  & 0.011583  & 10.879313 \\
\cline{2-5} & 3 & 0.008015  & 0.028096  & 14.047025 \\ \hline \hline
{SVM} & 1 & 0.011831  & 0.007723  & 10.060302 \\
\cline{2-5} & 2 & 0.012350  & {\bf 0.007703}    & {\bf 10.711573} \\
\cline{2-5} & 3 & 0.012412  & {\bf 0.007322}    & {\bf 11.551324} \\
\hline \hline {Naive forecast} & 1 & - & 0.077947
& 25.345352 \\ \cline{2-5} & 2 & - & 0.147725  & 34.918149 \\
\cline{2-5} & 3 & - & 0.207250  & 41.972591 \\ \hline

\end{tabular}
\caption{Experimental results on artificial time-series}
\label{res_artificial}
\end{center}
\end{table}

The optimal $l_{tw}$ for SVM was 3 for all predictions. Table
\ref{artificialPMBSI} shows the optimal settings found for PMBSI.
For $l_{pr} = 1$ when PMBSI outperformed linear SVM the optimal
length of the string map was shorter than the optimal time window
for SVM; in the remaining cases it was significantly longer.

\begin{table}
\begin{center}
\begin{tabular}[c]{|c|c|c|c|c|}
\hline $l_{pr}$ & $l_s$ & $Q$ & $\eta_1$ & $\eta_2$ \\ \hline 1 & 2
& 0.30 & 0.80 & -0.20 \\ \hline 2 & 5 & 0.10 & 0.80 & -0.60  \\
\hline 3 & 8 & 0.10 & 0.80 & -0.60  \\ \hline
\end{tabular}
\caption{Optimal PMBSI parameters} \label{artificialPMBSI}
\end{center}
\end{table}

\clearpage

\section{Prediction model based on the deviations from the closed string/pattern form (PMBCS)}
For the next trading strategy we want to define some real values of
the string sequences. Therefore we define the momentum which
acquired values from the interval $(0,1)$. The momentum $M$ is not
strictly invariant as in the previous model of the time series in
its basic definition. It is a trading strategy to find such place in
forex time-series market where $M$ is exactly invariant or almost
invariant and we can predict increasing or decreasing of prices with
higher efficiency. For example our predictor somewhere in
time-series has $55\%$ of efficiency to predict movement of price
but in the invariant place of our trading strategy where Eq. 26 is
almost invariant the efficiency of our predictor increased to
$80\%$. Therefore the idea to find invariant in time series plays a
crucial role in our trading strategy but one still needs to find an
appropriate expression for such prediction.

To study the deviations from the benchmark string sequence we define
momentum as
\begin{eqnarray}
&&M_{(l_s,m; Q,\varphi)} =\left(\frac{1}{l_s+1}
\sum_{h=0}^{l_s}\left| \frac{p(\tau +h)-p_{\rm min}(\tau)}{ p_{\rm
max}(\tau)-p_{\rm min}(\tau)} - \frac{1}{2} \left(1+
\cos\left[\frac{2\pi m h}{l_s+1}+\varphi\right]
\right)\right|^Q\right)^{1/Q}
\end{eqnarray} where
\begin{eqnarray}
p_{\rm stand}(\tau;h;l_s) &=& \frac{p(\tau+h)   - p_{\rm
min}(\tau;l_s)}{p_{\rm max}(\tau;l_s) - p_{\rm
min}(\tau;l_s)}\nonumber \,,\qquad  p_{\rm stand} \in
(0,1),\nonumber \end{eqnarray} and
\begin{eqnarray}
p_{\rm max}(\tau;h;l_s)=\max_{h\in  \{0,1,2,  \ldots, l_{\rm s}\}}
p(\tau + h) \,,\qquad p_{\rm min}(\tau;h;l_s)=\min_{h \in \{0,1,2,
\ldots, l_{\rm s}\}} p(\tau+ h), \nonumber
\end{eqnarray}
and $\varphi$ is a phasee of periodic function. The momentum defined
above takes the values from the interval $M_{(l_s,m;
Q,\varphi)}\in(0,1)$. Periodic function $\cos(\tilde{\varphi})$ in
definition of Eq. 26 could be substituted by other types of
mathematical functions. The results with different kind of functions
could be different.

\subsection{Elementary trading strategy based on the probability
density function of $M$} \noindent The purpose is to take advantage of it
whenever the market conditions are favourable. As in the previous
model we are detrending forex data into the one dimensional topological
object "strings" with different parameters. The trading strategy is
based on the description of rate curve intervals by one value called
moment of the string. These moments are statistically processed and
some interesting values of moments are found. The values directly
affect opening and closing of trade positions. Algorithm works in
two complement phases. First phase consists of looking for "good"
values of moments followed by second phase which uses results from
the first phase and opening/closing of trade positions occur.
Simultaneously the first phase is looking for new "good" values of
moments.

Risk is moderated by a number of allowed trades that algorithm can
open during certain period. Also it is moderated by two paramaters
which affect selection of suitable moments for trading. Maximum
number of trades is 10 per hour. Algorithm is tested on various
periods of historical date. The number and period of simultaneously
opened trades are all the time monitored.

First set of parameters describes the moment (simple scalar
function of several variables from the interval (0,1) ). First set
consists of these parameters: length of moment string (number of
ticks or time period), quotient or exponent of moment, frequency of
moment function, phase shift of moment function. Second set of
parameters controls trading strategy and consists of these
variables: maximum number of simultaneously opened trades, skewness
of moments distribution and sharpe ratio of closed trades. As soon as algorithm calculates the value of moment and finds out that the value is
"good", then it immediately carries out an appropriate command.

The risk of algorithm is governed by the second set of parameters and can
vary from zero (low risk but also low or zero number of trades) to
the boundary values controlled by the model parameters. These
boundary values are unlimited but could be easily affected by the
skewness and sharpe ratio. These parameters can limit loss to
certain value with accuracy $\pm2$ percent but also limit overall
profit significantly if low risk is desired.

Arbitrage opportunity taking advantage of the occurrence of
difference in distribution. Opportunity is measured by {\em
Kullback-Liebler} divergence
\begin{equation}
D_{\rm KL} = \sum_{j (bins)} \mbox{pdf}(M^{+}(j)) \,\mbox{log} \, \,
\left( \, \frac{\mbox{pdf}(M^{+}(j))}{\mbox{pdf}(M^{-}(j))}
\,\right)
\end{equation}
where larger $D_{\rm KL}$ means better opportunities ($D_{\rm
KL}>D_{\rm threshold}$) e.q. when $D_{KL} > D_{threshold}$ it means
the buying Euro against USD could be more profitably. Statistical
significance means the smaller the statistics accumulated into bins
$\mbox{pdf}(M^{+}(j))$, $\mbox{pdf}(M^{-}(j))$, the higher is the
risk ($M$ from selected range should be widespread).

More generally we can construct the series of $(l_s+1)$ price tics
[$p(\tau), p(\tau+1), \ldots, p(\tau+l_s)]$ which are transformed
into single representative real value $M(\tau+l_s)$. Nearly
stationary series of $M(\tau+l_s)$ yields statistics which can be
split into: branch where $M$ is linked with future {uptrend}/{
downtrend} and branch where $M$ is linked with future { profit}/{
loss} taking into account {transaction costs}. Accumulation of
$\mbox{pdf}(M_{\rm long}^{+-})$ means (profit+ / loss-) or
$\mbox{pdf}(M_{\rm short}^{+-})$ (profit+ / loss-). $M^+$ in Eq. 27
describes when Eq. 26 bring profit and $M^-$ loss.

As in the previous section the model was again tested for various
sets of free parameters $l_s$, $h$, $Q$, $\varphi$. This model can
make ``more-tick'' predictions (in tests it varies from 100 to 5000
ticks). Therefore it is much more successful than the previous
model. It is able to make final profit of around $160\%$ but this
huge profit precedes a fall down of $140\%$ of the initial state. It
is important to emphasize that all profits mentioned here and below
are achieved by using leverage (borrowing money) from 1 to 10. The
reason for leverage is the fact that the model could simultaneously
open up to 10 positions (one position means one trade i.e.\ one pair
of buy-sell transactions). If one decides not to use any leverage
the final profit decreases 10 times. On the other hand, with using
the leverage 1 to 20 the final profit doubles itself. Of course, the
use of higher leverages is riskier as also dropdowns are higher.
There is, for example, in Fig. 7 a dropdown circa $6\%$ around 600
trades. With the use of leverage 1 to 20 this dropdown rises to
$12\%$.

$128000$ combinations of model's parameters have been calculated.
Figures 7-10 describe some interesting cases of the prediction
behavior of the model with the transaction cost included on the
EUR/USD currency rate of forex market. Figures 7,8 describe the
model (one set of parameters) under condition that the fall
down must not be higher than $5\%$. The best profit achieved in this
case is $12\%$.

In order to sort out the best combinations of parameters it is helpful to use the
statistical quantity called Sharpe ratio. The Sharpe ratio is a
measure of the excess return per unit of risk in a trading strategy
and is defined as
\begin{equation}
S = \frac{E(R-R_f)}{\sigma},
\end{equation}
where $R$ is the asset return, $R_f$ is the return on a benchmark
asset (risk free), $E(R-R_f)$ is the mean value of the excess of the
asset return over the benchmark return, and $\sigma$ is the standard
deviation of the excess of the asset return. You mention the sharpe
ratio. The values of sharpe ratio for the best fit are e.g. for Fig.
10 it is value $1.896$ and for Fig. 11 it is value $1.953$, where as
a reference profit we choose bank with $5\%$ profit.

Figure 9 shows the case where the Sharpe ratio has the highest value
from all sets of the calculated parameters. One year profit is
around $26\%$ and the maximum loss is slightly over $5\%$. Figure 10
describes the case requiring high value of Sharpe
ratio and with the aim to gain profit over $50\%$.

There exist sufficiently enough cases with high Sharpe ratio which
leads to enhancement of the model to create self-education model.
This enhancement takes some ticks of data, finds out the best case
of parameters (high Sharpe ratio and also high profit) and starts
trading with these parameters for some period. Meanwhile, trading
with previously found parameters model is looking for a new best
combination of parameters. Figure 11 describes this self-education
model where parameters are not chosen and the model itself finds the
best one from the financial data and is subsequently looking for the
best values for the next trading strategy.

\begin{figure}
 \begin{center}
 \includegraphics[height=7cm]{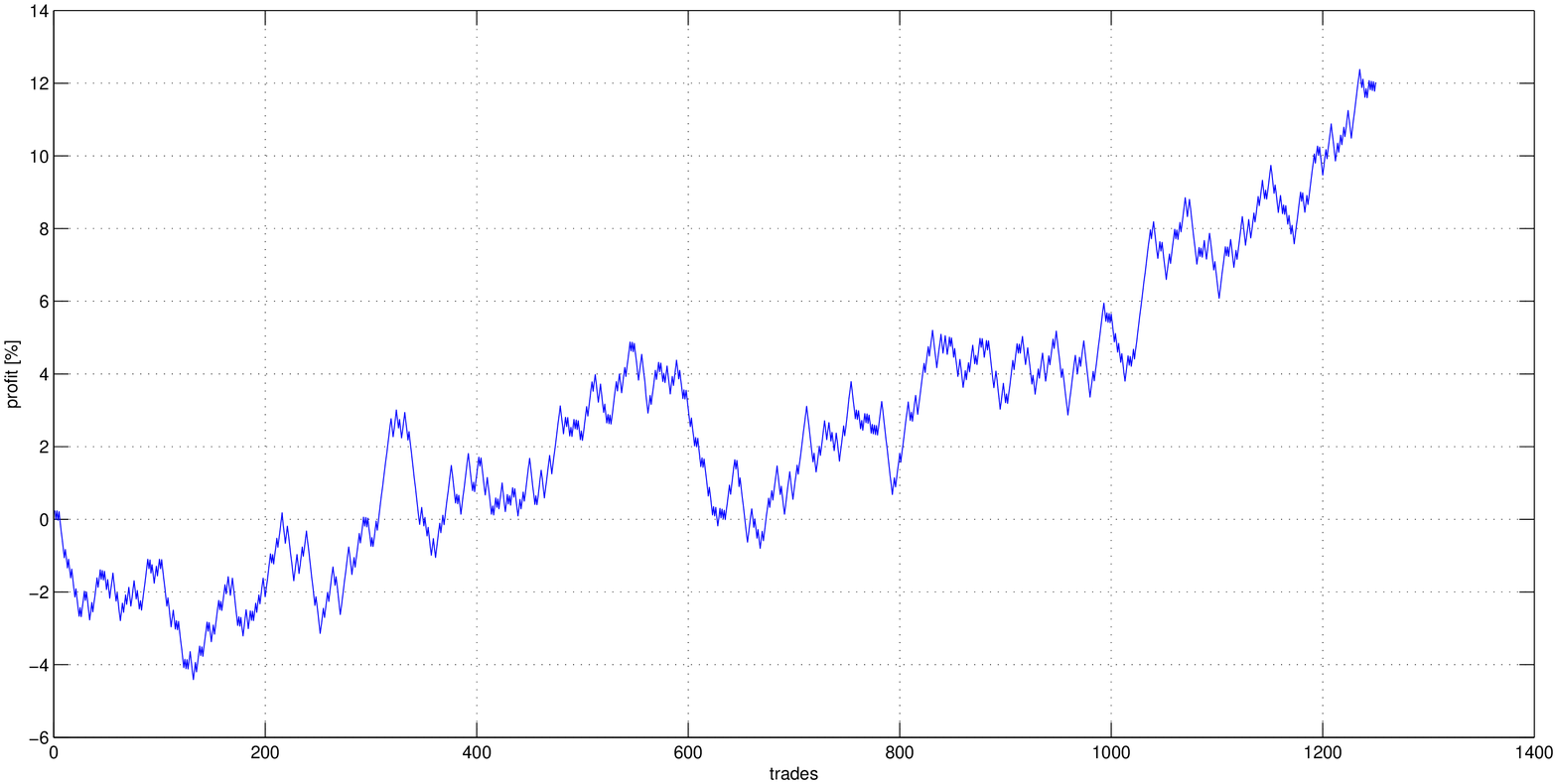}
 \caption{The profit of the model on the EUR/USD currency rate with transaction costs included dependence on trades for one year period.}
 \end{center}
 \end{figure}

\begin{figure}
 \begin{center}
 \includegraphics[height=7cm]{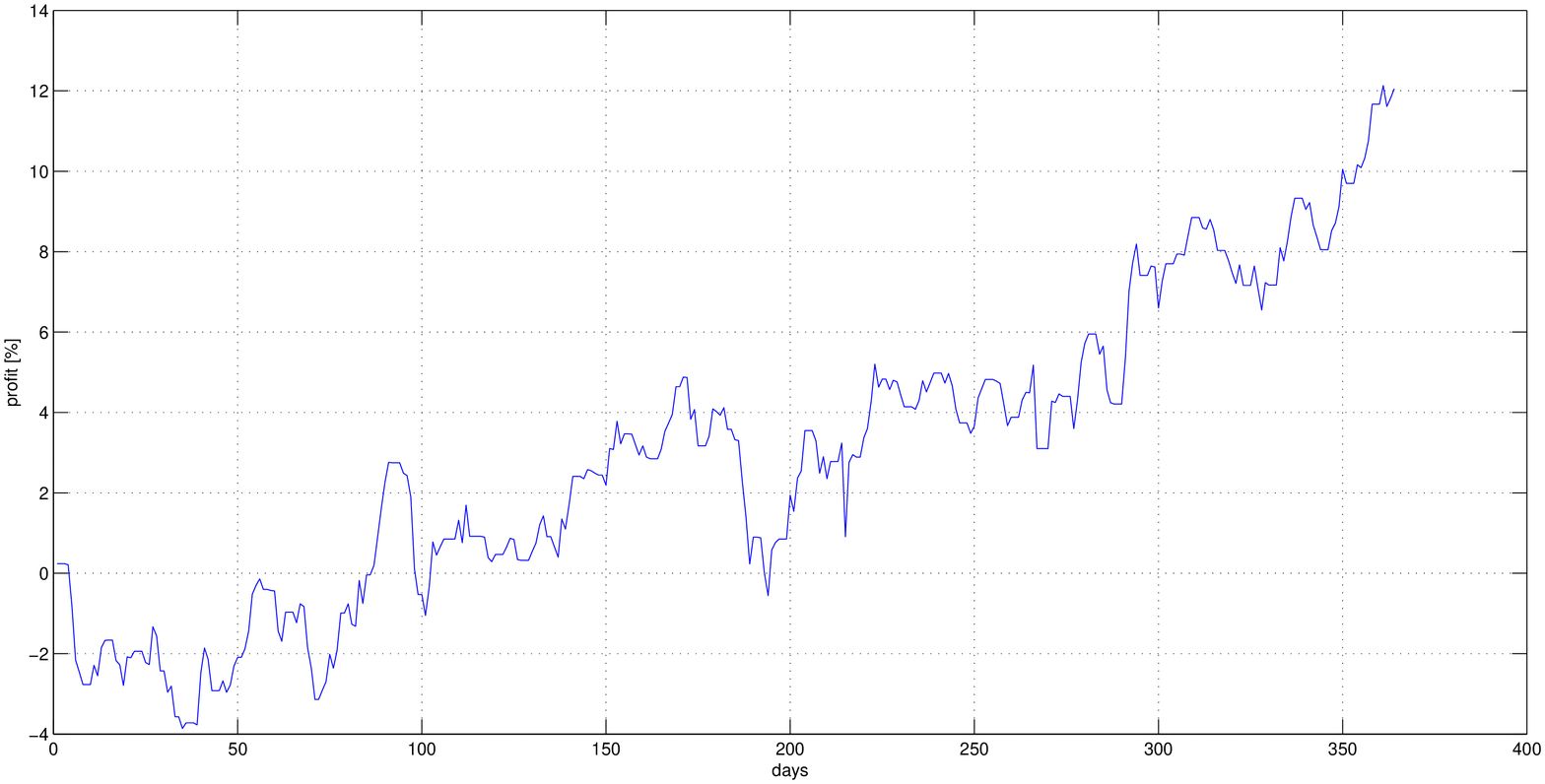}
 \caption{The profit of the model on the EUR/USD currency rate with transaction costs included dependence on days for one year period.}
 \end{center}
 \end{figure}

\begin{figure}
 \begin{center}
 \includegraphics[height=7cm]{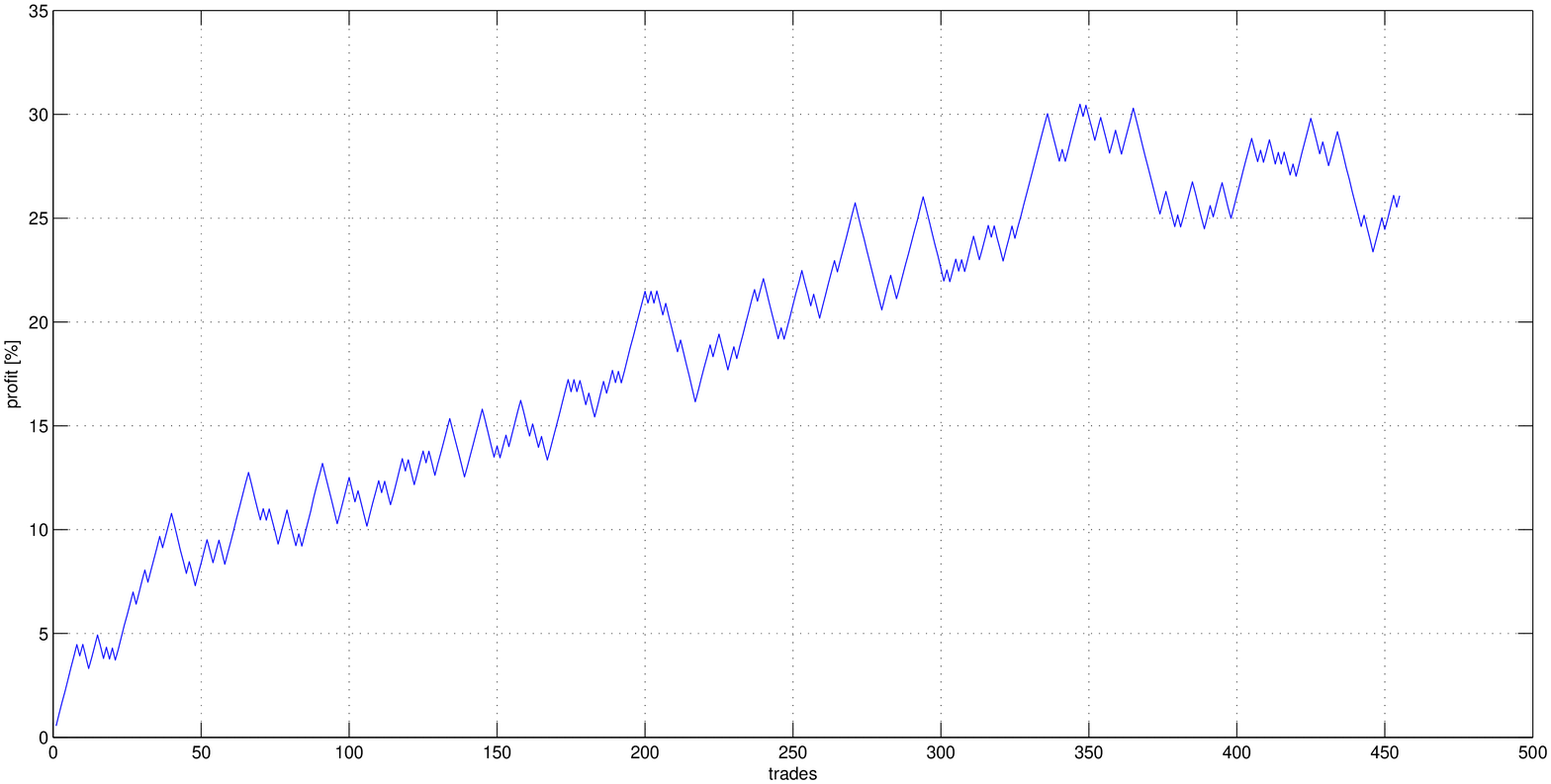}
 \caption{The profit of the model on the EUR/USD currency rate with transaction costs included dependence on trades for one year period.}
 \end{center}
 \end{figure}

\begin{figure}
 \begin{center}
 \includegraphics[height=7cm]{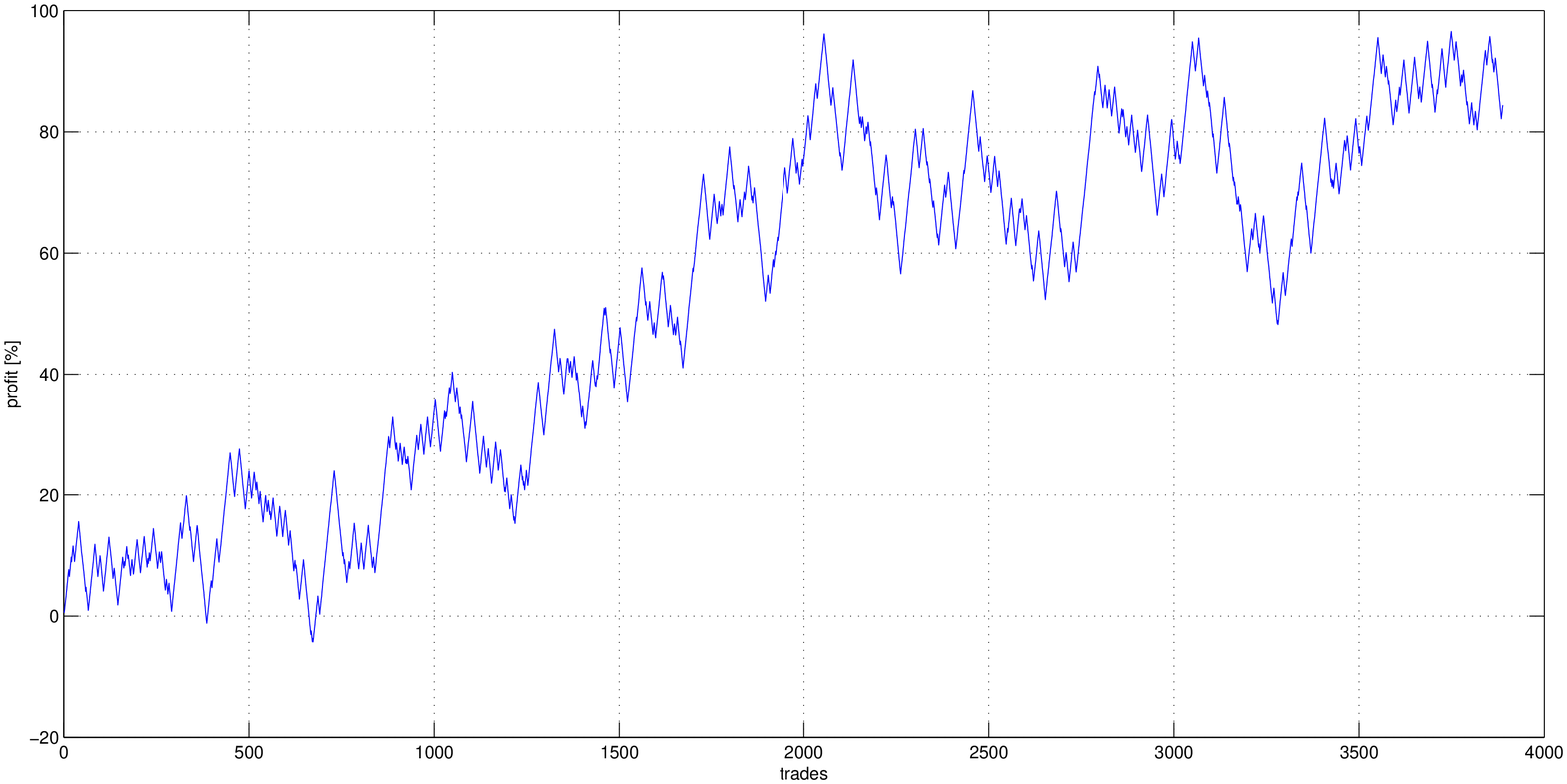}
 \caption{The profit of the model on the EUR/USD currency rate with transaction costs included dependence on trades for one year period.}
 \end{center}
 \end{figure}

\begin{figure}
 \begin{center}
 \includegraphics[height=7cm]{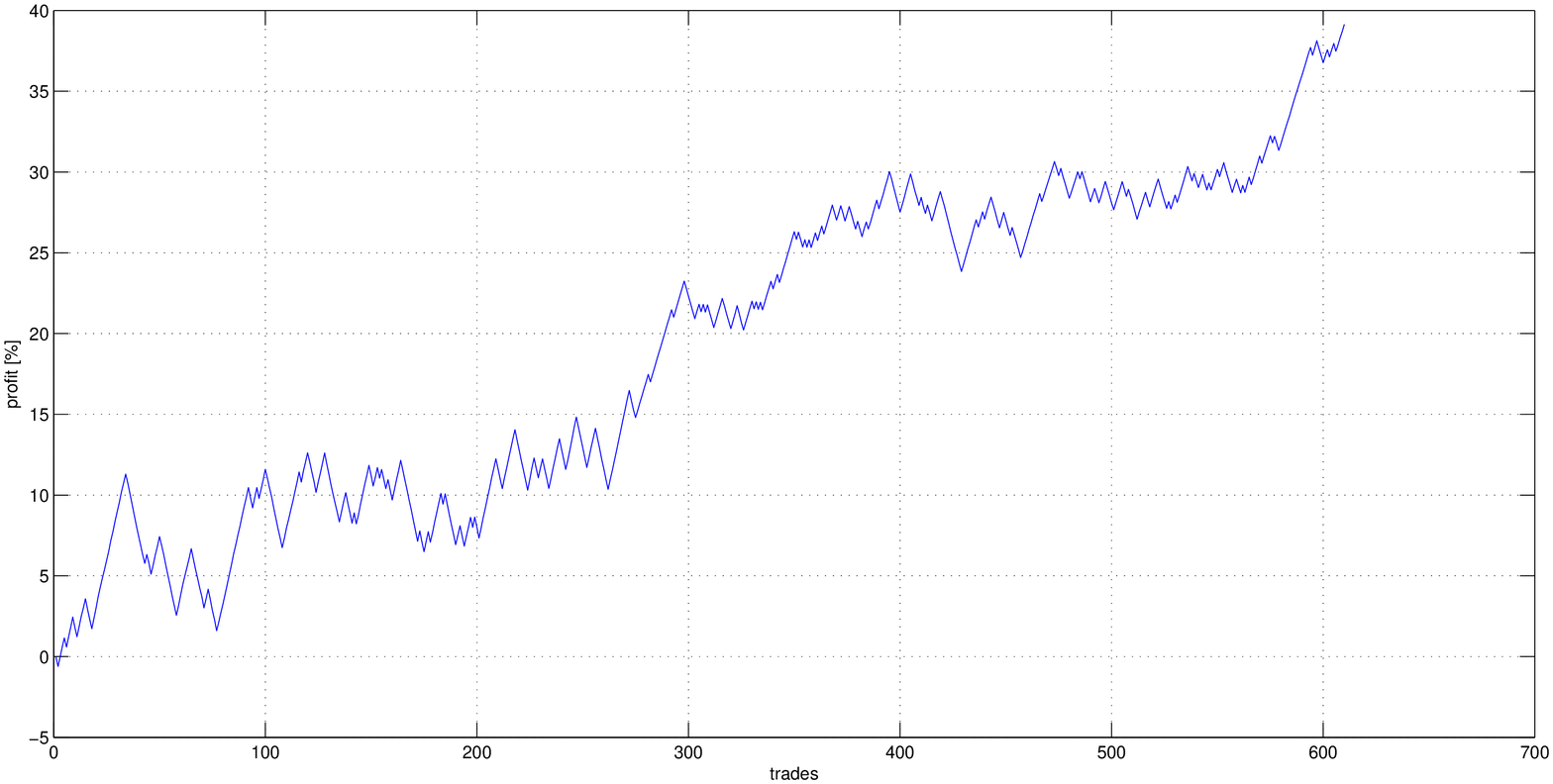}
 \caption{The profit of the self education model on the EUR/USD currency rate with transaction costs included dependence on trades for one year period.}
 \end{center}
 \end{figure}

\section{Conclusions}
The model of the strings allows one to manipulate with the
information stored along several extra dimensions. We started from
the theory of the 1-end-point and 2-end-point open string and
continued with partially compactified strings that satisfy the
Dirichlet and Neumann boundary conditions. We have $5$ free
parameters in our model. We have also tried out-of-sample tests, however, only using small data samples. We haven't encountered
"overfitting" due to the fact that parameters are stable enough within our
string theory approach to produce profit even if we slightly change
them. For all computations in the second model we are taking bid-offer
spreads into account. We are calculating with real values of
bid-offer spreads from historical data and it is dependent on where
we are simulating on Oanda or Icap etc. A number of trading per day
varies from $2$ to $15$ depending on fit strategy.

We have shown that the string theory may motivate the adoption of
the nonlinear techniques of the data analysis with a minimum impact
of justification parameters. The numerical study revealed
interesting fundamental statistical properties of the maps from the
data onto string-like objects. The main point here is that the
string map gives a geometric interpretation of the information value
of the data. The results led us to believe that our ideas and
methodology can contribute to the solution of the problem of the
robust portfolio selection.

We established two different string prediction models to predict the
behaviour of forex financial market. The first model PMBSI is based
on the correlation function as an invariant and the second one PMBCS
is an application based on the deviations from the closed
string/pattern form. We found the difference between these two
approaches. The first model cannot predict the behavior of the forex
market with good efficiency in comparison with the second one which,
moreover, is able to make relevant profit per year. From the results
described we can conclude that the invariant model as one step price
prediction is not sufficient for big dynamic changes of the current
prices on finance market. As can be seen in Figs. 3,4 when the
transaction costs are switched off the model has some tendency to
make a profit or at least preserve fortune. It means that it could also be useful but for other kind of data, where the dynamics of
changes are slower, e.g. for energetic~\cite{Toffolo} or
seismographic data~\cite{Stefansson} with longer periods of changes.
Finally the PBMSI in the form presented in this paper should be
applicable with good efficiency only to other kind of data with
smaller chaotic behaviour in comparison with financial data.

Moreover PMBSI is a method under development. Unlike SVM or ANN, at
this stage PMBSI does not require a training process optimizing a
large number of parameters. The experimental results indicate that
PMBSI can match or outperform SVM in one step ahead forecasts. Also,
it has been shown that finding optimal settings for PMBSI may be
difficult but the method's performance does not vary much for a wide
range of different settings. Besides the further test of PMBSI we
consider that fast methods for optimization of parameters must be
developed. Because of the character of the error surface we have
chosen to use evolutionary optimization as the method of choice.
After a fast and successful parameters' optimization method is
developed optimization of the weighting parameters (Eqs. 9,14) will
be included into the evolutionary process.

The profit per year from the second prediction model was obtained
from approximately 15 $\%$ and more depending on the parameter set
from the data we have chosen. This model is established efficiently
on finance market and could be useful to predict future prices for
the trading strategy. Of course the model still needs to be tested
further. With the flow of new financial data the model can be
more optimized and also, it could become resistant to crisis. The
presented models are universal and could also be used for
predictions of other kind of stochastic data. The self-educated
models presented in Fig. 11 are very useful because they are able to
find on their own the best parameter set from data, learn about the prices
and utilize these pieces of information for the next trading strategy.

\vskip 0.4cm ACKNOWLEDGEMENTS --- All calculations were done on the
computational cluster designed by the company FURT Solutions,
s.r.o., Kosice, Slovak Republic, http://www.furt.sk. For more
information about the progress of our new string application work to
financial market please follow the line
http://www.sors.com/people/pincak/.

\end{document}